\documentclass[conference]{IEEEtran}
\IEEEoverridecommandlockouts
\usepackage{cite}
\usepackage{amsmath,amssymb,amsfonts}
\usepackage{algorithmic}
\usepackage{graphicx}
\usepackage{textcomp}
\usepackage{xcolor}
\def\BibTeX{{\rm B\kern-.05em{\sc i\kern-.025em b}\kern-.08em
    T\kern-.1667em\lower.7ex\hbox{E}\kern-.125emX}}
    
\usepackage{booktabs} 
\usepackage{multirow}
\usepackage{booktabs}
\usepackage{graphicx}  
\usepackage{diagbox}
\usepackage[braket]{qcircuit} 
\usepackage{threeparttable} 
\usepackage{stfloats} 
\usepackage{enumitem} 
\usepackage{fancyheadings} 

\usepackage{times}
\usepackage{graphicx}
\usepackage{epsf}
\usepackage{verbatim}
\usepackage{url}
\usepackage{color}
\usepackage{alltt}

\newcommand{\Caption}{\caption}

\newcommand{\Comment}[1]{}

\newcommand{\SmallSpace}{\vspace*{-1.4ex}}

\begin{document}

\title{Some Size and Structure Metrics for \\Quantum Software
}

\author{\IEEEauthorblockN{Jianjun Zhao}
\IEEEauthorblockA{\textit{Kyushu University}\\
zhao@ait.kyushu-u.ac.jp}
}

\maketitle

\begin{abstract}
Quantum software plays a critical role in exploiting the full potential of quantum computing systems. As a result, it is drawing increasing attention recently. As research in quantum programming reaches maturity with a number of active research and practical products, software metric researchers need to focus on this new paradigm to evaluate it rigorously and quantitatively. As the first step, this paper proposes some basic metrics for quantum software, which mainly focus on measuring the size and structure of quantum software. These metrics are defined at different abstraction levels to represent various size and structure attributes in quantum software explicitly. The proposed metrics can be used to evaluate quantum software from various viewpoints.
\end{abstract}


\section{Introduction}
\label{sec:introduction}

Software metrics aim to measure the inherent complexity of software systems to predict the overall project cost and evaluate the quality and effectiveness of the design. Software metrics have many applications in software engineering tasks such as testing~\cite{hetzel1993making}, maintenance~\cite{gilb1988principles}, reengineering~\cite{arnold1992software}, reuse~\cite{poulin1996measuring}, and project management~\cite{fenton2014software,zuse2013framework}. 
Research for software measurement must adapt to the emergence of new software development methods, and metrics for new languages and design paradigms must be defined based on models relevant to these new paradigms ~\cite{bieman1996metric}. 

Quantum programming is the process of designing and building executable quantum computer programs to achieve a particular computing result~\cite{miszczak2012high,ying2016foundations}. A number of quantum programming approaches are available, for instance, Scaffold~\cite{abhari2012scaffold}, Qiskit~\cite{ibm2017qiskit}, Q\#~\cite{svore2018q}, ProjectQ~\cite{projectq2017projectq}, and Quipper\cite{green2013quipper}. A quantum program consists of blocks of code, each of which contains classical and quantum components. Quantum operations can be divided into {\it unitary} operations (reversible and preserve the norm of the operands), and {\it non-unitary} operations (not reversible and have probabilistic implementations). A quantum program executed on a quantum computer uses a quantum register of qubits to perform quantum operations and a classical register of classic bits to record the measurements of the qubits' states and apply quantum operators conditionally~\cite{cross2017open}. 

As research in quantum programming reaches maturity with a number of active research and practical products, software metrics researchers need to focus on this new paradigm to evaluate it in a rigorous and quantitative fashion~\cite{zhao2020quantum,piattini2020talavera}. 
However, although a large number of software metrics have been proposed for classical software~\cite{bieman1996metric,zhao1998assessing,fenton2014software,halstead1977elements,mccabe1976complexity,henry1981software,zuse2013framework}, few metrics have been proposed for quantum software. Further, due to quantum software's specific features, existing models and abstractions for classical software cannot be applied to model quantum software straightforwardly. 

As the first step towards evaluating quantum software, this paper proposes some basic metrics for quantum software, which mainly focus on measuring the size and structure of quantum software. These metrics are defined at different abstraction levels to represent various size and structure attributes in quantum software explicitly. The proposed metrics can be used to evaluate quantum software from various viewpoints.

The rest of this paper is organized as follows.  
Section~\ref{sec:background} reviews the fundamental terminology in quantum computing. Section~\ref{sec:size-metrics} and Section~\ref{sec:structure-metrics} propose some basic size and structure metrics for quantum software, respectively. Related work is discussed in Section~\ref{sec:work}, and concluding remarks are given in Section~\ref{sec:conclusion}. 

\section{Background}
\label{sec:background}

This section briefly introduces some basics of quantum mechanics, which form the basis of quantum computing~\cite{nielsen2002quantum}. 

\subsection{Quantum Bit}
\label{subsec:qubit}

A classical bit is a binary unit of information used in classical computation. It can take two possible values, 0 or 1. A quantum bit (or qubit) is different from the classical bit in that its state is theoretically represented by a linear combination of two bases in the quantum state space (represented by a column vector of length 2). We can define two qubits $|0\rangle$ and $|1\rangle$, which can be described as

\begin{center}
$|0\rangle = \begin{bmatrix}1 \\0 \end{bmatrix}$ and $|1\rangle = \begin{bmatrix}0 \\1 \end{bmatrix}$
\end{center}

\noindent
Qubits $|0\rangle$ and $|1\rangle$ are the computational basis state of the qubit. In other words, they are a set of the basis of quantum state space.

Any qubit $|e\rangle$ can be expressed as a linear combination of two basis as $|e\rangle = \alpha|0\rangle + \beta|1\rangle$, where $\alpha$ and $\beta$ are complex numbers, and $|\alpha|^2+|\beta|^2 = 1$. This restriction is also called {\it normalization conditions}.

\subsection{Quantum Gate and Circuit}
\label{subsec:q-gate}

Just as a logic gate in a digital circuit that can modify the state of a bit, a quantum gate can change the state of a qubit. A quantum gate can have only one input and one output (transition of a single quantum state), or it can have multiple inputs and multiple outputs (transition of multiple quantum states). The number of inputs and outputs should be equal because the operators need to be reversible, which means no information can be lost in quantum computing. 

\subsubsection{NOT Gate}
\label{subsubsec:not}

The NOT gate works on a single qubit. It can exchange the coefficients of two basis vectors:

\begin{center}
$NOT(\alpha |0\rangle + \beta |1\rangle) = \alpha |1\rangle + \beta |0\rangle$
\end{center}

\noindent
The quantum NOT gate is an extension of the NOT gate in classical digital circuits. 

A single input-output quantum gate can be represented by a $2\ \times\ 2$ matrix. The state of a quantum state after passing through the quantum gate is determined by the value of the quantum state vector left-multiplied by the quantum gate matrix. The quantum gate matrix corresponding to the NOT gate is
$X = \begin{bmatrix}0&1\\1&0\end{bmatrix}$
%
Therefore, the result of a qubit passing a NOT gate can be denoted as

\begin{center}
$X \begin{bmatrix}\alpha \\ \beta \end{bmatrix} = \begin{bmatrix}0&1\\1&0\end{bmatrix} \begin{bmatrix}\alpha \\ \beta \end{bmatrix} = \begin{bmatrix}\beta \\ \alpha \end{bmatrix}.$
\end{center}

\subsubsection{Hadamard Gate}
\label{subsubsec:hadamard}

The Hadamard gate also works on a single qubit, which can decompose existing quantum states according to its coefficients as follows.

\begin{center}
$H(\alpha |0\rangle + \beta |1\rangle) = \frac{\alpha + \beta}{\sqrt{2}}|0\rangle + \frac{\alpha - \beta}{\sqrt{2}}|1\rangle$ 
\end{center}

This can be represented by a matrix as: 
$$H = \frac{\sqrt{2}}{2}\begin{bmatrix}1&1\\1&-1\end{bmatrix}$$

\noindent
Although Hadamard gate is not directly related to the AND and OR gates in classical digital circuits, it has important applications in many quantum computing algorithms. Interested readers can try to prove that after applying the Hadamard gate twice in a row, the quantum state will return to its original state. This behavior is consistent with the NOT gate.


\subsubsection{Controlled NOT Gate}
\label{subsubsec:controlled}

Computer programs are full of conditional judgment statements: if so, what to do, otherwise, do something else. In quantum computing, we also expect that the state of one qubit can be changed by another qubit, which requires a quantum gate with multiple inputs and outputs. The following is the controlled-NOT gate (CNOT gate). It has two inputs and two outputs. If the input and output are taken as a whole, this state can be expressed by 
\begin{center}
$\alpha |00\rangle + \beta |01\rangle + \gamma |10\rangle + \theta |11\rangle$
\end{center}
\noindent
where $|00\rangle$, $|01\rangle$, $|10\rangle$, $|11\rangle$ are column vectors of length 4, which can be generated by concatenating $|0\rangle$ and $|1\rangle$. This state also needs to satisfy the normalization conditions, that is $|\alpha|^2 + |\beta|^2 + |\gamma|^2 + |\theta|^2 = 1$.

The CNOT gate is two-qubit operation, where the first qubit is usually referred to as the control qubit and the second qubit as the target qubit. When the control qubit is in state $|0\rangle$, it leaves the target qubit unchanged, and when the control qubit is in state $|1\rangle$, it leaves the control qubit unchanged and performs a Pauli-X gate on the target qubit. It can be expressed in mathematical formulas as follows:

\vspace*{1mm}
\noindent
{\footnotesize 
$CNOT(\alpha |00\rangle + \beta |01\rangle + \gamma |10\rangle + \theta |11\rangle) = \alpha |00\rangle + \beta |01\rangle + \gamma |11\rangle + \theta |10\rangle$
}

\vspace*{2mm}
\noindent
The action of the CNOT gate can be represented by the matrix:
\begin{center}
{\footnotesize
$X = \begin{bmatrix}1&0&0&0\\0&1&0&0\\0&0&0&1\\0&0&1&0\end{bmatrix}$
}
\end{center}

\subsubsection{Quantum Circuit}
\label{subsubsec:q-circuit}

Quantum circuits, also known as quantum logic circuits, are the most commonly used general-purpose quantum computing models, which represent circuits that operate on qubits under an abstract concept. 
A quantum circuit is a collection of interconnected quantum gates. The actual structure of quantum circuits, the number and the type of gates, and the interconnection scheme, are all determined by the unitary transformation $U$, performed by the circuit. The result of a quantum circuit can be read out through quantum measurements.

\subsection{Quantum Programming}
\label{subsec:programming}
Quantum programming is the process of designing and building executable quantum computer programs to achieve a particular computing result~\cite{miszczak2012high,ying2016foundations}. A quantum program consists of blocks of code, each of which contains classical and quantum components. Quantum operations can be divided into {\it unitary} operations (reversible and preserve the norm of the operands), and {\it non-unitary} operations (not reversible and have probabilistic implementations). A quantum program executed on a quantum computer uses a quantum register of qubits to perform quantum operations and a classical register of classic bits to record the measurements of the qubits' states and apply quantum operators conditionally~\cite{cross2017open}. Therefore, a typical quantum program usually consists of two types of instructions (or statements). One is called {\it classical instructions} that operate on the state of classical bits and apply conditional statements. Another is called {\it quantum instructions} that operate on the state of qubits and measure the qubit values.

\section{Basic Size Metrics for Quantum Software}
\label{sec:size-metrics}

Software size represents one of the most significant internal attributes of a software product~\cite{fenton2014software}. Many software effort estimation models~\cite{albrecht1983software,jones2008applied,boehm1995cost} use some size metrics as their main effort driver to estimate the software development effort. In this section, we introduce some basic metrics that can be used to measure the size of quantum software from different levels: {\it code}, {\it design}, and {\it specification}.  

\subsection{Code Size}
Program code is an integral component of the software. Such code includes source code, intermediate code, and even executable code. Here we only focus on the source code of quantum software and discuss how some widely used size metrics for classical software can be extended to measure the size of quantum software.

\subsubsection{Lines-of-Code (LOC)}
As we know, the most commonly used metric of source code program length is the number of lines of code (LOC). The LOC metrics can be used to predict the amount of effort required to develop a program and can also be used to estimate the programming productivity or maintainability after the software product is completed.
Similarly, for quantum software, we may also consider some LOC metrics to measure the internal attribute of the software, specifically related to the quantum attributes. 
Let $Q$ be a quantum program, some basic LOC metrics for $Q$ can be defined as follows: 

\begin{itemize}
    \item $\varphi_{1}$: number of LOC in $Q$
\end{itemize}


\noindent
As we observed, $\varphi_{1}$ is a general metric, which consists of both classical and quantum statements in $Q$, that can be used to measure the total size of a quantum program. We can also define some metrics to especially measure the size related to the quantum aspects in $Q$ as follows. 

\begin{itemize}
    \item $\varphi_{2}$: number of LOC related to quantum gate operations
    \item $\varphi_{3}$: number of LOC related to quantum measurements
\end{itemize}

Based on the above metrics, we can finally derive a general size metric to quantify the total size of the quantum-related aspects in $Q$ as $\varphi_{4} = \varphi_{3} + \varphi_{2}$. 

Moreover, we can even consider the number of qubits or gates used in $Q$ to obtain the following size metrics.

\begin{itemize}
    \item $\varphi_{5}$: total number of qubits used in $Q$
    \item $\varphi_{6}$: total number of unique quantum gates used in $Q$
\end{itemize}

\begin{figure}[t]
\footnotesize{
  \begin{alltt}
  (1)  simulator = \textbf{Aer.get_backend}('qasm_simulator')
  (2)  qreg = \textbf{QuantumRegister}(2)
  (3)  creg = \textbf{ClassicalRegister}(2)
  (4)  circuit = \textbf{QuantumCircuit}(qreg, creg)
  (5)  circuit.\textbf{x}(q[1])
  (6)  circuit.\textbf{h}(q[0])
  (7)  circuit.\textbf{h}(q[1])
  (8)  circuit.\textbf{cx}(q[0],q[1])
  (9)  circuit.\textbf{h}(q[0])
  (10) circuit.\textbf{h}(q[1])
  (11) \textbf{for} i \textbf{in} range(2):
  (12)     circuit.\textbf{measure}(q[1-i], c[i])
  (13) job = \textbf{execute}(circuit, simulator)
  (14) result = job.\textbf{result}()
  (15) counts = result.\textbf{get_counts}(circuit)
  (16) \textbf{print}(counts)
  \end{alltt}
}
\Caption{\label{fig:qiskit-example}A simple quantum program in Qiskit.}
\end{figure}

Figure~\ref{fig:qiskit-example} shows a simple Qiskit program. From the above definitions, we can calculate the value of each LOC metric for the program to obtain the following results:
\begin{center}
$\varphi_{1} = 16$, $\varphi_{2} = 6$, $\varphi_{3} = 1$, $\varphi_{4} = 7$, $\varphi_{5} = 2$, and $\varphi_{6} = 3$.
\end{center}

\subsubsection{Halstead's Software Science}
Halstead~\cite{halstead1977elements} introduces some size metrics for classical software. According to Halstead, "a computer program that implements an algorithm is considered to be a collection of tokens that can be divided into operators or operands." Halstead observes that software metrics should reflect how algorithms are implemented in different programming languages and independent of the platform and language used. These metrics should be computable statically from the source code of the program.

Let $P$ be a classical program which is a collection of tokens, classified as either {\it operators} or {\it operands}, then some basic numbers for these tokens in $P$ can be defined as follows:

\begin{itemize}
    \item $\mu_{1}$: number of unique operators
    \item $\mu_{2}$: number of unique operands
    \item $N_{1}$: total occurrences of operators
    \item $N_{2}$: total occurrences of operands
\end{itemize}

\noindent
Based on these numbers, several metrics could be defined further to quantify various aspects of classical software.  

\begin{itemize}
    \item The length of $P$: \hspace*{1mm}$N = N_{1} + N_{2}$
    \item The vocabulary of $P$: \hspace*{1mm}$\mu = \mu_{1} + \mu_{2}$ 
    \item The estimated length of $P$: \hspace*{1mm}$N_{E} = \mu_{1} \log_2 \mu_{1} + \mu_{2} \log_2 \mu_{2}$ 
    \item The volume of $P$: \hspace*{1mm}$V = N \times \log_2 \mu$
    \item The difficulty: \hspace*{1mm}$D = \frac{\mu_{1}}{2} \times \frac{N_{2}}{\mu_{2}}$ 
    \item The effort required to generate $P$: \hspace*{1mm}$E = D \times V$
\end{itemize}

Let $Q$ be a quantum program that consists of both classical and quantum statements, and we can define some basic numbers in $Q$ as well.

\begin{itemize}
    \item $\eta_{1}$: number of unique classical and quantum operators
    \item $\eta_{2}$: number of unique classical and quantum operands
    \item $M_{1}$: total occurrences of classical and quantum operators
    \item $M_{2}$: total occurrences of classical and quantum operands
\end{itemize}

\noindent
We can define some metrics for quantum software by extending the classical Halstead's metrics based on these numbers. These metrics can be used to quantify various aspects of quantum software.

\begin{itemize}
    \item The length of $Q$: \hspace*{1mm}$M = M_{1} + M_{2}$
    \item The vocabulary of $Q$: \hspace*{1mm}$\eta = \eta_{1} + \eta_{2}$ 
    \item The estimated length of $Q$: \hspace*{1mm}$M_{E} = \eta_{1} \log_2 \eta_{1} + \eta_{2} \log_2 \eta_{2}$ 
    \item The volume of $Q$: \hspace*{1mm}$V_{\scriptscriptstyle Q} = M \times \log_2 \eta$
    \item The difficulty of $Q$: \hspace*{1mm}$D_{\scriptscriptstyle Q} = \frac{\eta_{1}}{2} \times \frac{M_{2}}{\eta_{2}}$ 
    \item The effort of $Q$: \hspace*{1mm}$E_{\scriptscriptstyle Q} = D_{\scriptscriptstyle Q} \times V_{\scriptscriptstyle Q}$
\end{itemize}


\subsection{Design Size}
Quantum software design is a process to transform user requirements into a suitable form that helps the programmer in quantum software implementation (coding). As classical software design [75], a quantum software design may also involve two stages: {\it architectural design} and {\it detailed design}. Like measuring the size of quantum software at the code level, we can measure the size of a quantum software design at the architectural level as well. 

\subsubsection{Architectural Design Size} 
Architectural design defines a collection of quantum software components, their functions, and their interactions (interfaces) to establish a framework for developing a quantum software system. The software architecture of the system defines its high-level structure, revealing its overall organization as a set of interacting components~\cite{perry1992foundations,mary1996software}. A well-defined architecture allows an engineer to reason about system properties at a high level of abstraction. We can measure the size of a quantum architectural design based on formal architectural specifications and architectural patterns.

Architectural description languages (ADLs)~\cite{clements1996survey} are formal languages representing the architecture of a software system. Using ADLs, software architects can specify the system's functionality using components and the interaction between components using connectors. 

Like classical ADLs, we can use a quantum architectural description language (qADL), an extension of a classical ADL to formally specify the architectures of a quantum software system, to support the architectural-level design of quantum systems. A qADL can specify a quantum system's architecture with those mechanisms, including specifications of classical components and connectors between classical components, specifications of quantum components and connectors between quantum components, and specifications of connectors between classical and quantum components. 

Let $S$ be the formal architectural specification of a quantum software architecture, which consists of a group of components and the connectors between components, and we can define some general architectural-level size metrics for $S$ as follows:

\begin{itemize}
\item $\gamma_{1}$: number of lines of architectural specifications
\item $\gamma_{2}$: number of components and connectors
\end{itemize}

\noindent
We can also define some size metrics that specifically quantify the size of the quantum aspects in $S$ from different viewpoints.

\begin{itemize}
\item $\gamma_{3}$: number of quantum components
\item $\gamma_{4}$: number of connectors between quantum components
\item $\gamma_{5}$: number of connectors between a quantum component and a classical component
\end{itemize}

Based on the above metrics, we can finally derive a general size metric to quantify the total size of the quantum-related aspects in $S$ as $\gamma_{6} = \gamma_{3} + \gamma_{4} + \gamma_{5}$

\subsubsection{Detailed Design Size}
Detailed design refines and expends the architectural design to describe the internals of the quantum software components (the algorithms, processing logic, data structures, and data definitions).

Recently, some researches have been carried out for studying design patterns for quantum systems~\cite{weigold2020data,leymann2019towards}. We can define some design size metrics for quantum systems based on these quantum patterns.

Let $D$ be a design for a quantum system, and we can define some general size metrics in terms of design patterns in $D$ 

\begin{itemize}
\item $\delta_{1}$: number of unique design patterns used in $D$
\item $\delta_{2}$: number of design pattern realizations for each pattern type
\end{itemize}

Moreover, we can also define some size metrics that are specific to quantify the quantum attributes in terms of quantum design patterns in $D$ as follows:

\begin{itemize}
\item $\delta_{3}$: number of unique quantum design patterns used in $D$
\item $\delta_{4}$: number of quantum design pattern realizations for each pattern type
\end{itemize}

\subsection{Specification Size}
The Unified Modeling Language (UML)~\cite{uml2009version} is a general-purpose, well-known modeling language in the field of classical software engineering. It provides a standard way to visualize the design of the classical software development life cycle. 
Recently, P\'{e}rez-Delgado and Perez-Gonzalez~\cite{Perez-Delgado2020quantum} present an extension of UML called {\it Q-UML} to model quantum software systems. 
The extension covers two types of UML diagrams: {\it class diagram} and {\it sequence diagram}. In this work, they claim that, in addition to the classical elements in a classical UML, the Q-UML should contain those elements related to the quantum aspects, such as quantum classes, quantum elements (quantum variables and quantum operations), quantum supremacy, quantum aggregation, and quantum communications.  

Similar to the classical UML, Q-UML can serve as a base for defining some size metrics for a quantum software system at the specification level. An example of some metrics, which can be used to quantify the quantum aspects of the system, can be defined as follows.

\begin{itemize}
\item $\theta_{1}$: number of quantum classes (objects)
\item $\theta_{2}$: number of quantum elements (quantum variables or quantum operations)
\item $\theta_{3}$: number of quantum interfaces
\item $\theta_{4}$: number of quantum attributes
\item $\theta_{5}$: number of all quantum methods
\end{itemize}

\section{Basic Structure Metrics for Quantum Software}
\label{sec:structure-metrics}

Structural analysis of programs is an essential component of software development and software evaluation efforts. Structure metrics try to take into account the complexity of individual modules and the interactions between modules in a product or system and quantify such interactions. Many approaches in structure metrics have been proposed. Some good examples include McCabe's cyclomatic complexity metric~\cite{mccabe1976complexity} and the information flow metric by Henry and Kafura~\cite{henry1981software}. In this section, we discuss how these classical software metrics can be extended to evaluate quantum software.

\subsection{McCabe's Complexity Metric}
\label{subsec:QSLC}

A well-known metric for measuring the complexity of a classical program's control structure is the cyclomatic complexity metric by McCabe~\cite{mccabe1976complexity}. 

For a program $P$, the cyclomatic complexity metric $V(G)$ of $P$ is defined as the number of linearly independent paths through a control-flow graph (CFG) of $P$ and can be computed as follows:

\begin{center}
$V(G) = E - N + 2$
\end{center}

\noindent
where $G$ is the CFG of $P$ to be measured, $N$ is the number of nodes of $G$, each representing a statement in $P$, and $E$ is the number of edges of $G$, each presenting the flow of control from one statement to another in $P$.  

Similarly, we can measure the complexity of a quantum program's control structure by defining an extended cyclomatic complexity metric for quantum software. The definition of such a metric can be based on the quantum control flow graph (QCFG), which is an extension of the classical CFG to represent a quantum program. 

Let $G_{\scriptscriptstyle Q}$ be the QCFG of a quantum program $Q$ to be measured, the cyclomatic complexity metric $V(G_{\scriptscriptstyle Q})$ of $Q$ can be defined as follows.

\begin{center}
$V(G_{\scriptscriptstyle Q}) = E_{\scriptscriptstyle Q} - N_{\scriptscriptstyle Q} + 2$
\end{center}

\noindent
where $N_{\scriptscriptstyle Q}$ is the number of nodes of $G_{\scriptscriptstyle Q}$ and $E_{\scriptscriptstyle Q}$ is the number of edges of $G_{\scriptscriptstyle Q}$. 

Similar to classical CFG, each node $n\in N_{\scriptscriptstyle Q}$ represents either a classical (or quantum related) statement, and each $e \in E_{\scriptscriptstyle Q}$ represents the flow of control from one statement to another in $Q$.  
The QCFG of $Q$ can be constructed based on its AST, which can be derived through the static analysis on $Q$. 

\begin{figure}[t]
\centerline{\includegraphics[width=0.8\linewidth]{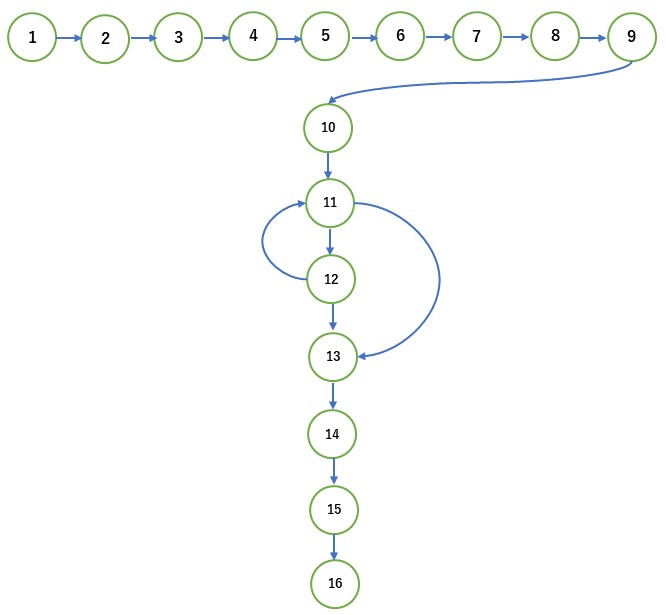}}
\caption{The quantum control-flow graph of the Qiskit program in Figure~\ref{fig:qiskit-example}.}
\label{fig:qcfg}
\end{figure}

As an example, Figure~\ref{fig:qcfg} shows the QCFG of the Qiskit program in Figure~\ref{fig:qiskit-example}. Based on the QCFG, we can calculate that the cyclomatic complexity metric of the program is 17 - 16 + 2 = 3.    

\subsection{Henry and Kafura's information flow Metric}

Perhaps the most common design structure metrics are known as information flow metrics, which deal with the complexity of a system by observing the flow of information among system components or modules. Among these metrics, one well-known approach is Henry and Kafura's information flow metric~\cite{henry1981software}, which measures the total level of information flow between individual modules and the rest of a system.  
This metric considers the complexity of a software module, which could be affected by the following two factors.

\begin{itemize}
\item The complexity of the module code itself.
\item The complexity due to the module's connections to its environment. 
\end{itemize}

The effect of the first factor has been included through the LOC (Lines Of Code) metric. For the quantification of the second factor, Henry and Kafura have defined two terms: {\it fan-in} and {\it fan-out}.

\begin{itemize}
    \item The {\it fan-in} of a module $M$ is the number of local flows into $M$, plus the number of data structures from which $M$ retrieves information.
    \item The {\it fan-out} of a module $M$ is the number of local flows from $M$, plus the number of data structures which $M$ updates.
\end{itemize}

Based on these concepts, Henry and Kafura define their information flow metric value as the module length multiplied by the square of fan-in multiplied by fan-out.

\begin{center}
$IF(M) = length(M) \times (F_{in}(M) \times F_{out}(M))^2$
\end{center}

\noindent
where $length(M)$ is the number of programming language statements in $M$, and $F_{in}$ and $F_{out}$ are the fan-in and fan-out of $M$, respectively. 

A quantum software system is usually composed of classical and quantum modules. These modules are not independent, and therefore may have some interactions among them, which may lead to some information flows, from a classical/quantum module to a classical/quantum module. To measure the complexity of a quantum system by observing the flow of information among the system modules, we can extend Henry and Kafura's information flow metric to the domain of quantum systems.

Let $M_{\scriptscriptstyle Q}$ be a classical/quantum module of a quantum system, $length(M_{\scriptscriptstyle Q})$ be the number of quantum programming language statements in $M_{\scriptscriptstyle Q}$, and $F_{in}$ and $F_{out}$ be the fan-in and fan-out of $M_{\scriptscriptstyle Q}$, respectively, we can define a structure metric based on the information flow to measure the complexity as a module of fan-in and fan-out in the system as follows.

\begin{center}
$IF(M_{\scriptscriptstyle Q}) = length(M_{\scriptscriptstyle Q}) \times (F_{in}(M_{\scriptscriptstyle Q}) \times F_{out}(M_{\scriptscriptstyle Q}))^2$
\end{center}

\section{Related Work}
\label{sec:work}
Although a large body of research in software metrics has been proposed for classical software~\cite{bieman1996metric,zhao1998assessing,fenton2014software,halstead1977elements,mccabe1976complexity,henry1981software}, few metrics have been proposed for quantum software. 
To the best of our knowledge, our work is the first to propose some size and structure metrics for quantum software from various viewpoints. 

Several metrics have been proposed to measure the performance of quantum computer systems. One is called the {\it Total Quantum Factor (TQF)} by Sete {et al.}~\cite{sete2016functional}. TQF can provide rough estimates of the quantum circuit's size (circuit width times circuit depth) that can be run before the processor's performance decoheres. Another metric used to measure quantum computing systems' performance is called {\it Quantum Volume}~\cite{bishop2017quantum}. Salm {\it et al.}~\cite{salm2020criterion} propose a metric to measure the performance of a gate-based quantum computer, which can be used to determine if a quantum circuit is successfully executable on a given gate-based quantum computer. They also discuss how the quantum system, noise, and errors affect the metric and how it can be refined for precise estimation.


Piattini {\it et al.}~\cite{piattini2020talavera} present the Talavera Manifesto for quantum software engineering and programming, which collects some principles and commitments about the field of quantum software engineering and programming. Among these principles and commitments, one is ``{\it QSE assures the quality of quantum software}," which mentioned that new metrics for quantum software should be developed. However, they do not define any concrete metric regarding quantum software measurement in the manifesto. 

Sicilia {\it et al.}~\cite{sicilia2020source} present a preliminary study on the structure of the source code of quantum software, using initially the same metrics typically used in classical software. Their study focuses on the module structure and use of quantum gates in the libraries of Microsoft's quantum development platform QDK (Quantum Development Kit) that uses a specific language Q\#. However, they do not propose any software metric that can be used to quantify the attributes of quantum software. 

\section{Concluding Remarks}
\label{sec:conclusion}
This paper has proposed some basic metrics for quantum software, which mainly focus on measuring the size and structure of quantum software. Some of these metrics are the extensions of their classical counterparts, while the others are specifically designed to quantify the quantum features in quantum software. These metrics are defined at different abstraction levels to represent various size and structure attributes in quantum software explicitly. The proposed metrics can be used to evaluate quantum software from various viewpoints.

We are now implementing a metrics collection tool for quantum programming language Qiskit to automatically derive the results of proposed metrics from Qiskit programs. The next step for us is to perform some experiments and collect data for evaluation. We hope a primary evaluation of these metrics will be available soon.

\section*{Acknowledgment}
We are grateful to the anonymous reviewers for their suggestions to improve the paper and Hidenori Ooka for his valuable discussions.

\bibliographystyle{IEEEtran}
\bibliography{IEEEabrv,qse-bibliography}

\end{document}